\documentclass[twocolumn,useAMS,usenatbib]{revtex4}
\pdfoutput=1
\usepackage{graphics,epsfig}
\usepackage[dvipsnames]{color}
\usepackage[]{inputenc,amssymb}

\def \be{\begin{equation}}
\def \ee{\end{equation}}
\def \bea{\begin{eqnarray}}
\def \eea{\end{eqnarray}}

\def\simle{\lower 2pt \hbox {$\buildrel < \over {\scriptstyle \sim }$}}
\def\simge{\lower 2pt \hbox {$\buildrel > \over {\scriptstyle \sim }$}}

\def\AJ{{\it AJ} }
\def\ARAA{{\it ARA\&A} }
\def\ApJ{{\it ApJ} }
\def\ApJL{{\it ApJL} }

\def\AA{{\it A\&A} }

\def\AAL{{\it Astron. \& Astroph. Letters} }

\def\MNRAS{{\it MNRAS} }
\def\Nature{{\it Nature} }
\def\NewAR{{\it New Astron. Rev.} }

\def\PRD{{\it Phys. Rev. D.} }
\def\PRL{{\it Phys. Rev. Lett.} }

\def\Science{{\it Science}Ê}

\begin{document}
\title{Dark energy as stimulated emission of gravitons from a background brane }
\author{Peter L. Biermann}
\email{plbiermann@mpifr-bonn.mpg.de}
\affiliation{Department of Physics and Astronomy, The University
of Alabama, Box 870324, Tuscaloosa, AL 35487-0324, USA}
\affiliation{MPI for Radioastronomy, Bonn, Germany}
\affiliation{Karlsruhe Institute of Technology (KIT) - Institut f{\"u}r Kernphysik, Germany}
\affiliation{Department of Physics \& Astronomy, University of Bonn, Germany}
\author{Benjamin C. Harms}
\email{bharms@bama.ua.edu}
\affiliation{Department of Physics and Astronomy, The University
of Alabama, Box 870324, Tuscaloosa, AL 35487-0324, USA}
%
%


\label{firstpage}

\begin{abstract}
The idea that dark energy is gravitational waves may explain its strength and its time-evolution provided that the additional energy comes from a background.  A possible concept is that dark energy is the ensemble of coherent bursts (solitons) of gravitational waves originally produced by stimulated emission when the first generation of super-massive black holes was formed.  These solitons get their initial energy as well as keep up their energy density throughout the evolution of the universe by stimulating emission from a background brane.  We model this process by working out this energy transfer in a Boltzmann equation approach.   The transit of these gravitational wave solitons may be detectable.  Key tests include pulsar timing, clock jitter and the radio and neutrino backgrounds. \vspace{3mm}

\end{abstract}

%
\maketitle

\section{Introduction}

Dark energy was originally detected as accelerated expansion seen in the distance scale for supernovae of type Ia (\cite{Schmidtetal98,Riessetal98,Perlmutteretal99}; for a review see \cite{Friemanetal08}).  Many suggestions have been made about what dark energy is, what its strength is, what its time evolution is, and what possible further observational results might be.

The idea that dark energy is gravitational waves produced by stimulated emission from a background brane may explain its strength and its time-evolution.  One possible concept is that dark energy is the ensemble of coherent bursts (solitons) of gravitational waves originally produced when the first generation of super-massive black holes was formed \cite{CarameteBiermann10}; the energy density of such solitons would suffice within the uncertainties.  Stimulated emission from a background brane provides the initial energy of these solitons as well as keeps up their energy density throughout the evolution of the universe \cite{BH1,BH2,BH3}.  The process of this energy transfer modifies the Raychaudhuri-equation (eq.\ref{FRW2}) such that it becomes consistent with an equation of state $P_{DE}\,=\,\rho_{DE}\,c^2/3$.    Our model of the background metric resembles the Randall-Sundrum ideas \cite{RSa,RSb} but  is time-dependent, and describes the energy flow from the background (strong-gravity) brane to our world (weak-gravity) brane. Planck data suggest that dark energy has increased in strength over cosmic time (\cite{Planck2013XVI}), as predicted by our model.  Gravitational waves were far below today's dark energy at the epoch of early nucleosynthesis and of the formation of the microwave background ripples (as summarized in \cite{LigoVirgo09}), both much earlier than the likely formation epoch of the first generation of super-massive black holes.  Our model is also consistent with early star formation,  as we argue below.  The transit of the gravitational wave solitons postulated here may be detectable.  
We discuss the predictions briefly below and elsewhere \cite{BH3,Zeuthen12}.  We focus on the Boltzmann equation approach, working out the energy transfer from the strong gravity background in stimulated emission.

Inspired by Bekenstein's \cite{Bekenstein73} considerations we posit:  When the first generation of super-massive black holes was formed, each produced a coherent burst of soliton-like gravitational waves which combine to give a total energy of order
\be
\rho_{DE} \; \simeq \; \frac{1}{2} \, N_{BH,0} \, M_{BH} \, c^2 \, (1 + z_{\star})^3 \; .
\label{blackholeeq}
\ee
We justify this expression below. 

In the following we also call this an ensemble of soliton waves, or shell fronts.  $N_{BH,0}$ is the original comoving density of super-massive black holes, formed at redshift $z_{\star}$. Today super-massive black holes have a density of $10^{- 1.7 \pm 0.4} \; {\rm Mpc^{-3}}$ above $M_{BH} = 3 \cdot 10^{6} \, {\rm M_{\odot}}$ \cite{CarameteBiermann10}; assuming that they grow by merging, and allowing for statistical and systematic errors, an original comoving density of $N_{BH,0} \, = 1 \; {\rm Mpc^{-3}}$ seems possible. This comoving density is the density black holes had at the beginning, so transposed to today without change in their numbers per comoving volume.  The data suggest that there was a generation of first super-massive black holes with a mass between $M_{BH} \; \sim \; 10^{6} \, M_{\odot}$ and $M_{BH} \; \sim \; 10^{7} \, M_{\odot}$.  The original black hole mass may be $\sim \; 3 \cdot 10^{6} \, M_{\odot}$ considering (i) the black hole mass function \cite{GreeneBarthHo06,CarameteBiermann10}, (ii) the instability of super-massive stars \cite{AppenzellerFricke72a,AppenzellerFricke72b} in an agglomeration growth of such stars \cite{Spitzer69,Sanders70}, and (iii) the observed black hole in our Galactic Center (e.g. \cite{Eckartetal05}).  The redshift of creation $z_{\star}$ may be large, as formation of massive stars may begin at redshift of about 80 \cite{BiermannKusenko06}.  Redshifts $z_{\star}$ from about 30 to 50 allow a quantitative interpretation of  the data of dark energy.  At the original density of black holes adopted here for reference of $N_{BH,0} \, = 1 \; {\rm Mpc^{-3}}$ redshift 50 is consistent with the mass of $M_{BH} \; \sim \; 3 \cdot 10^{6} \, M_{\odot}$, and redshift 30 would imply $M_{BH} \; \sim \; 10^{7} \, M_{\odot}$, in either case to make the estimate in eq.(\ref{blackholeeq}) consistent with dark energy today.

What is the motivation for considering gravitational waves?  Bekenstein \cite{Bekenstein73} wrote about the entropy of the universe:  ``... we must regard black hole entropy as a genuine contribution to the entropy content of the universe".  However, entropy is also information, and information must have a carrier. A natural suggestion is that this carrier is gravitational waves, with an energy commensurate with the black hole scale.  This suggestion is consistent with the fact that cosmological black holes are not in thermodynamic equilibrium, and therefore the entropy associated with such black holes should be described by statistical mechanics as advocated in Harms and Leblanc (\cite{har92,har93}).  This speculation immediately gives
\be
\frac{S}{k_{B}} \; = N_{GW,0} \; = \; \; 4 \pi {\left(\frac{M_{BH}}{m_{Pl}}\right)}^2 \;\; , {\rm \; for \; zero \; spin} \; ,
\ee
where $N_{GW,0}$ is the number of gravitons at the formation of the black hole.  For $M_{BH} \, = \, 3 \cdot 10^{6} \, {\rm M_{\odot}}$ this is $N_{GW,0} \; \simeq \; 10^{90}$.  $M_{BH}$ is the original mass of the black hole, $m_{Pl}$ is the Planck mass, $c$ is the speed of light, and $G_{N}$ is Newton's constant of gravity.  $E_{GW}$ is the average graviton energy given by 
\be
E_{GW} \, = \, \frac{1}{8 \pi} \; \frac{\hbar c^3}{G_N M_{BH}} \; = \; \frac{c^2}{8 \pi} \, \frac{m_{Pl}^2}{M_{BH}}\; .
\ee
This gives a graviton  energy of  $E_{GW} \, \simeq \, 10^{-30}  \, {\rm erg}$ for this black hole  mass.  The entire energy content then is
\be
N_{GW,0} \, E_{GW}  \; = \; \frac{1}{2} \, M_{BH} \, c^2 \; .
\label{InjEq}
\ee
which, multiplied by the original density, $N_{BH,0}$, of black holes, reproduces eq.(\ref{blackholeeq}), at the original redshift $z_{\star}$.
We picture this as a coherent burst of gravitational waves, or a soliton wave, ejected at formation of the black hole.  It is clear from the considerations above that we are not using the weak-field approximation.  We derive this injection term from a Boltzmann equation as described in Section III.


\section{Background model}

In our model for the background, which has some similarity to the Randall-Sundrum \cite{RSa,RSb} ideas. In our model energy transfer occurs between the two branes: The condition is such that stimulated emission happens only if the phase-space density is locally larger on our brane than on the strong brane (see eq.\ref{pvol}).  We identify a possible local metric to describe a 5D world with a 4D strong gravity brane and our 4D world weak-gravity brane.
\begin{eqnarray}
ds^2 = -e^{A_1(t,u)}\,c^2\,dt^2+e^{A_2(t,u)}\,dx_i\,dx^i+e^{A_3(t,u)}\,du^2   \, ,
\label{metric} 
 \end{eqnarray}
 \noindent where $i\,=\,1,2,3$, and $u$ is the coordinate in the fifth dimension.   The three functions $A_i(t,u)$ can be determined from the Einstein equations for the model in which the five-dimensional universe is described by a perfect fluid.  In our model the five-dimensional universe behaves as a Friedmann-Robertson-Walker (FRW) universe (see \cite{Gergely03,Gergely08} for a much broader discussion).  However, the projections of the metric in eq.(\ref{metric}) onto the strong- and weak-gravity branes result in modifications of the four-dimensional  FRW equations on those branes.  We define $\rho_{DE}(t,u)$ as the dark energy density, $\rho(t,u)$ as the total energy density, $P(t,u)$ as the total pressure, and we use the equation of state
 \bea
 P_{DE}(t,u)\,=\,\frac{1}{3} \, \rho_{DE}(t,u) \, c^2
 \label{Eq6}
 \eea
 and then check the FRW equations for consistency.
 We begin by utilizing the five-dimensional conservation equation
 \bea 
 \nabla_{\mu}T^{\mu\nu}\,=\,0\, ,
 \eea 
 to determine the functions $A_1(t,u)$ and $A_3(t,u)$. These functions  are found to be for $\nu\,=\,4$ and $\nu\,=\,0$ respectively,
 \bea
 A_1(t,u)\,=\,\ln  \left( -1/2\,\ln  \left( {\it \rho_{DE}} \left( t,u \right) 
 \right) + F_1(t)  \right)\, ,
 \eea
 and
\bea
&&A_3(t,u)= -2\,\ln  \left( {\it \rho_{DE}} \left( t,u \right)  \right)  \label{Eq9}\\
&& -3\,\ln 
 \left( \ln  \left( {\it \rho_{DE}} \left( t,u \right)  \right) -2\,F_1(t) \right)
 -3\,{\it A_2} \left( t,u \right) +{\it F} \left( u
 \right) \, ,\nonumber
 \eea
 where $F(u)$ is an arbitrary function and $0\,\le u\,\le l_{SB}$. Here $l_{SB}$ is by magnitude a few hundred Planck lengths in order to match the inferred ratio of $10^{123}$ between the energy densities between the branes ($10^{123} \, \simeq \, e^{283}$). On our brane $A_1(t,0)= 0$ and $A_2(t,0) = \ln(R(t)^2)$. The factor $R(t)$ is scaled to the epoch when the black hole creation rate is the largest.  On the weak-gravity brane the function $A_3(t,0)$ is essentially zero, while on the strong-gravity brane $A_3(t,l_{SB})$ is large.  On the weak-gravity brane $\rho_{DE}(t,0)$ is observed to be a constant and the function $A_2(t,0)$ describes the adiabatic expansion of this brane.  The arbitrary function $F_1(t)$ is constant since $\rho_{DE}(t,0)$ is constant on the weak-gravity brane.  The function $A_3(t,u)$ describes the contraction of the coordinate $u$.

The Friedmann-Robertson-Walker form of the Einstein equations on the weak-gravity brane must be 
\bea
\left(H(t)\right)^2\,=\,\left(\frac{\dot{R(t)}}{R(t)}\right)^2\,=\,\frac{8\,\pi\,G_N}{3}\,\rho(t,0)
\label{FRW1}
\eea
and
\bea
\frac{\ddot{R}(t)}{R(t)}&=&-\frac{4\,\pi\,G_N}{3}\,\left(\rho(t,0)\,+ 3\,\frac{P(t,0)}{c^2}\right)\nonumber\\
&&+\,\frac{16\,\pi\,G_N}{3}\,\rho_{DE}(t,0)\,+\frac{4\,\pi\,G_N\,S_{inj}}{3\,H(t)}\, ,
\label{FRW2}
\eea
where the rate of change of the energy density used is
\bea
\dot{\rho}(t,0)=&-&3\left(\rho(t,0)+\frac{P(t,0)}{c^2}\right)H(t)\,\nonumber\\
&+&\,4\,\rho_{DE}(t,0)\, H(t)\,+\,S_{inj}\, .
\label{Econservation}
\eea 
These two equations can be derived using Lemaitre's argument \cite{LEM}.  In these expressions $4\,\rho_{DE}(t,0)\,H(t)$ is the rate of dark energy creation on our brane due to stimulated emission from the strong-brane and $S_{inj}$ is the rate of new dark energy creation due to black hole formation or due to the mergers of black holes. These equations replace the traditional Friedmann equations in the case of an explicit energy transfer.   We emphasize that in the second equation, eq.\ref{FRW2}, the term ${16\,\pi\,G_N}\,\rho_{DE}(t,0)/3$ corresponds to the continuous energy transfer by stimulated emission on the basis of existing dark energy (the additional term $({4\,\pi\,G_N\,S_{inj}})/({3\,H(t)})$ describes new formation of dark energy). This allows a different equation of state for exactly the same cosmological observations, as now for $P_{DE} = \rho_{DE}c^2 /3$ the modified equation \ref{FRW2} becomes identical to the canonical version of this equation for $P_{DE} = - \rho_{DE} c^2$.  The corresponding set of equations for the strong-brane are
\bea 
H_{SB}(t)^2 &=& \left(\frac{\dot{R}_{SB}(t)}{R_{SB}(t)}\right)^2 \nonumber\\
&=&\frac{8\,\pi\,G_N}{3}\left(\rho_{DE}(t,l_{SB})-\Lambda_{SB}\right)
\eea
where $\Lambda_{SB}$ is the cosmological constant on the strong brane at the beginning of the epoch of black hole formation and
\bea
&&\frac{\ddot{R}_{SB}(t)}{R_{SB}(t)}\,=\,-\frac{16\,\pi\,G_N}{3}\,\rho_{DE}(t,0)\,\frac{H(t)}{H_{SB}(t)}\\
&+&\frac{8\,\pi\,G_N}{3}\,\left(\,3\,\rho_{DE}(t,l_{SB})-\Lambda_{SB}\right)
-\frac{4\,\pi\,G_N\,\,S_{inj}}{3\,H_{SB}(t)}\, , \nonumber
\eea
The rate of change of the energy density on the strong-brane is correspondingly 
\bea
&\dot{\rho}_{DE}(t,l_{SB})=3\left(\rho_{DE}(t,l_{SB})+\frac{1}{c^2} P_{DE}(t,l_{SB})\right)H_{SB}(t)\nonumber\\
&-\,4\,\rho_{DE}(t,0)\,H(t)\,-\,S_{inj}\, .
\label{rhodot}
\eea
 \noindent so on the strong brane all is completely dominated by dark energy, in our concept gravitational waves. For epochs before the energy transfer started    $\rho_{DE}(t,0)=0$, and $S_{inj}=0$, and we can set $\rho_{DE}(t,l_{SB})\,=\,\Lambda_{SB}$ for $t \; < \; t(z = z_{\star})$.
 
The metric in eq.(\ref{metric}) describes a weak brane for our world which is expanding with time, and a strong brane which is contracting with time, albeit very slowly for the latter brane.  The 5D-cosmological constant measured on the weak brane is 

{\large$\Omega_{weak}\,=\,-\,{\left({\frac{\tau_{Pl}}{\tau_{H}}}\right)}^{2} \; \simeq \, 10^{-122.8}$}.  

The expressions in eq.\ref{Eq6} through eq.\ref{Eq9} are valid for all $t$ and $u$. For eqs.(\ref{FRW1}-\ref{rhodot}) to hold, we require that the sum of all the terms involving derivatives of $A_1(t,u),\, A_2(t,u),\,  {\rm and \; \; } A_3(t,u)$ with respect to $u$ all vanish on the two branes, separately in each of the Einstein equations; one can verify that i) this requirement is self-consistent, and ii) that all such terms are second order, so either the product of two first derivatives, or a second derivative.  So, from the Einstein equations with an energy-momentum tensor for a perfect fluid in five dimensions we obtain FRW equations which are valid for all $t$ and $u$.  We use the FRW equations on the two branes as given above as a consistency check of the behavior of the metric tensor elements.

In our model the gravitons at high energy in the background obey a Planck-like distribution with Planck temperature.  The energy density on the strong brane is the Planck energy divided by the Planck volume, which is a factor of order $10^{123}$ higher than the energy density on our brane.  The strong brane is stable against collapse, since given a Planck spectrum for any wavelength $\lambda$ the free-fall time scale $\tau_{ff}$ is always either equal or longer than the pressure wave time scale $\tau_{s}$.

{$\tau_{ff} \, = \, \tau_{Pl} {\left(\frac{\lambda}{l_{Pl}}\right)}^{3/2} \ge \, \tau_{Pl} {\left(\frac{\lambda}{l_{Pl}}\right)} \, = \, \tau_{s}$}.

\section{Energy transfer by stimulated emission}

In the following we use the particle-wave duality for gravitons at high energy, which thus associates a wavenumber $\vec{k}/\hbar$ and a  corresponding length-scale $\lambda$ to each spatial direction.  

\subsection{Direct derivation from Boltzmann equation}

The distribution function ${\mathcal N}(k,t)$ is the distribution of occupied allowed states for gravitons with momenta $k =|\vec{k}| $ and $p = |\vec{p}|$ at time $t$ on the shell and satisfies the equation (see, for example, \citep{bern,Kompaneets57,Zeldovich75})
\bea
&\left(\frac{\partial}{\partial t}-\frac{\dot{R}(t)}{R(t)}\, k\,\frac{\partial}{\partial k}\right)\,{\mathcal  {N}}(k,t) =\cr & \frac{1}{k}\int\,\frac{d^3\, k'}{(2\pi)^3\,2\, k'}\int\,\frac{c d^3\, p}{(2\pi)^3\,2\, E(p)}\int\,\frac{c d^3\,p'}{(2\pi)^3\,2\, E(p')}\int\,\frac{d^3\, k''}{(2\pi)^3\,2\, k''}\nonumber\\
\cr & \gamma^2 \, |M|^2\,(2\,\pi)^4\delta^4(K+Q-K'-K''-Q')\,\nonumber\\
\cr & (2 \, \pi)^6\,\delta^3(\vec{k'}-\vec{k''}) \,\delta^3(\vec{k}-\vec{k'})\, \nonumber\\
\cr & \left[g_{b}(p',t)\, (1+{\mathcal  {N}}(k,t)) \, \left( ({\mathcal  {N}}(k',t)+1)\,({\mathcal  {N}}(k'',t)+1) - 1\right) \,  \right.\nonumber\\
\cr &- \left. {\mathcal {N}}(k,t)\, g_{b}(p,t)\,(1+{\mathcal {N}}(k',t))\, (1+{\mathcal {N}}(k'',t))\right] \; .
\label{Boltzmann}
\eea
where $K = (k, \vec{k})$, $\gamma \, = \, k_{ref, 1}^3$,  and $Q = (E, \vec{p})$. $k_{ref, 1}$ is a reference momentum to be determined below.
The $\delta$-functions, $\delta^3(\vec{k'}-\vec{k''})$ and $\delta^3(\vec{k}-\vec{k'})$, have been inserted to impose coherence of the outgoing gravitons.  $|M|^2$ is the matrix element squared for the quadrupole emission of a graviton of 4-momentum $k''$, and has the dimensions of $(momentum)^{-3} \; (time)^{-1}$.  We will assume that this matrix element squared has the same dependence as that derived in \cite{BoughnRothman05}, which is proportional to $k^5$.  $g_b(p,t)$ is the occupation number distribution of the background particle sea.  $R(t) = (1+z_{\star})/(1+z) $ is the scale factor for an expanding universe.  The following analysis is done in the observer frame.

The Boltzmann equation for $\mathcal{N}(k,t)$ to lowest order in the expansion of the 4-dimensional $\delta$-function is
\be 
\left(\frac{\partial}{\partial t}-\frac{\dot{R}(t)}{R(t)}\, k\,\frac{\partial}{\partial k}\right)\,{\mathcal  {N}}(k,t) \simeq \frac{+\kappa}{k}\,{\mathcal{N}}(k,t)\,({\mathcal{N}}(k,t)+1) \, ,
\label{Bolt}
\ee 
where the factor $\kappa$ is given, after integration over all the $\delta$-functions, by 
%
\be 
\kappa\,=\,\frac{k_{ref, 2}^2 \, H(z)}{2^4 \pi k_{BH}} \, {|\overline{M}|^2} \, \ln\{\frac{k_{BH+}}{k}\} \, .
\label{kap}
\ee 
In the equation above $k_{BH+}$ is the maximum momentum at which stimulated emission of gravitons occurs, just above the momentum of the peak of $\mathcal  {N}(k,t)$.  Since the log-term in eq.\ref{kap} varies very slowly over the range of $k$ of interest we approximate this term with a constant and set $\beta \; = \; \{\ln(k_{BH+}/k)\}/(2^4 \, \pi)$.  $|M|^2$ is related to $|\overline{M}|^2$ by extracting the factors $(k/k_{ref,1})^3/k_{ref, 1}^3$, $H(z)$. 
The integration over the three $\delta$-functions in eq.\ref{Boltzmann} results in a factor of $k^{-4}$ in the phase-space expression.  For later convenience we choose to define $|\overline{M}|^2$ such that 3 of the 4 inverse powers are cancelled by the $k^5$ factor in $|M|^2$.  The factor $k_{ref, 1}^3$ then cancels one remaining earlier factor $\gamma$, so that this term disappears from now on for the expressions.  We choose these factors to make $|\overline{M}|^2$ dimensionless. A threshold function of $2 \, g_b\,\mathcal{N}/(g_b +\mathcal{N})$ arises from the normalized interaction between the gravitons on the background brane and our brane; this function connects to the strong brane only if $\mathcal{N}\; > \; g_b$, which is the condition for stimulated emission (see eq.\ref{pvol}).  We will specify these scaling terms for the momentum $k_{ref,1}$, $k_{ref,2}$, and $k_{BH}$ further down.  We note that the integral converges, since $d^3 p$ can be written as $4 \pi p^2 d p$, and the term $p^2 c^2$ cancels $E(p)^2$ in the denominator. We will show below that for stimulated emission from the background requires that locally in phase-space $g_b \, < \, \mathcal{N}(k,t)$, so that the integral is solvable, as $|\overline{M}|^2$ is a function of $k$ only.  We adopt $k_{ref,2} = m_{Pl} c$.  We will show below that this is consistent.

Next we redefine $(k/k_{ref,2})^2 \, {\overline{\kappa}} \; = \; \kappa$ to extract the $k$-dependence from $|\overline{M}|^2$, which we will justify below.

Making the change of variables $k\,=\,\tilde{k}/R(t)$,  eq.\ref{Bolt} can be written as 
\be 
\frac{\partial\mathcal{N}(\tilde{k},t)}{\partial t}\simeq\,\frac{+{H(z) \, \tilde{k} \,\beta} }{k_{BH} \, R(t)}\,\mathcal{N}(\tilde{k},t)\,(\mathcal{N}(\tilde{k},t)+1) \, .
\ee
In terms of the frequency of the wave at emission this equation is
\be 
\frac{\partial\mathcal{N}(\nu_0,t)}{\partial t}\simeq\,\frac{+{H(z) \, h\,\nu_0 \, \beta}}{k_{BH} \, R(t) \, c} \, \mathcal{N}(\nu_0,t)\,(\mathcal{N}(\nu_0,t)+1) \, .
\label{nu0}
\ee
Introducing the dimensionless variables
\be 
x\,=\,\frac{h\,\nu_0}{k_B\,T_{g0}}\, , \hspace{8mm}{\rm and}\hspace{8mm} y\,=\,\int_0^t\frac{k_B T_{g 0} \,H(z) \beta}{k_{BH} \, c \, R(t)}\,dt'\, ,
\ee
where { $k_{BH} c \, = \, k_B T_{g 0} = m_{Pl} c^2 \frac{m_{Pl}}{8\,\pi\,M_{BH}}$}.
eq.(\ref{nu0}) becomes
\be 
\frac{\partial\mathcal{N}}{\partial y}\,\simeq\,+x \, \mathcal{N}(x,y)\,(\mathcal{N}(x,y)+1) \, .
\ee
The solution of this equation is 
\be 
\mathcal{N}(x,y)\,=\,\frac{1}{e^{x(a-y)+b}-1}\, ,
\label{sol}
\ee
where $a$ and $b$ are constants, to be determined later.   This distribution (eq.\ref{sol}) is Planck-like with a time-dependent normalized temperature $1/a$.  

The rate at which energy is created can be calculated from the expression for ${\mathcal{N}}$ in eq.(\ref{nu0}).  The rate of energy creation per existing graviton (of the total number $N_{GW,0}\,R(t)^4$) is 
\bea 
&&\frac{d<E>}{dt} = \frac{\beta H(z)}{ R(t)} \\ && \int x^3 \, h \nu_{0} \, {\mathcal{N}}(\nu_0,t) \, ({\mathcal{N}}(\nu_0,t) +1) \, d  x\, .\nonumber
\eea
The total rate of energy creation is 
\be 
\frac{d<E_T>}{dt}\,=\,N_{GW,0}\,R(t)^3 \, k_{BH} \, c \, H(z) \, \beta \, A \; ,
\label{dEdt}
\ee 
where 
\be
A = \int x^4 \, {\mathcal{N}}(x,t) \, ({\mathcal{N}}(x,t) +1) \, d  x
\ee


  Inserting all these constants into the integral for $y$ demonstrates that $y$ approaches a constant for the redshift $z_{\star}$ being large, and integrating down to today or even into the future, when $y$ approaches a constant of $\beta \, << \, 1$.  Since $a$ is equivalent to an inverse temperature, it seems natural based on the scaling of the spectrum, that this quantity $a$ should be of order unity. Without loss of generality we can set $b \, = \, 0$.  This integral strongly depends on the exact value of $a - y$, and is of order 30 for $a - y \simeq 1$, and $b$ approaching zero.   
 

The matrix element $|\overline{M}|$ does not evolve with time, and scales as momentum

\be 
|\overline{M}|\,=\,\epsilon_M \, {\left(\frac{k  }{m_{Pl} c}\right)} .
\ee
Above we have used $\epsilon_M = 1$; we now generalize and allow $\epsilon_M$ to be different from unity.  Writing the term $|\overline{M}|$ in this way suggests, that we have succeeded condensing the interaction between the gravitons on our brane and the gravitons on the background brane down to a fundamental coupling constant.  This behavior is quite consistent with common expectations, that the gravitational coupling strongly increases with energy to approach the other three coupling constants at near Planck energies.

This allows the expression for { $\frac{d<E_T>}{dt}$} to be consistent with the observed energy density under the condition that $A \, \beta  \, \epsilon_M = 3$.  For the constant $a \, = \, 1$ above, $\beta$ of order 0.1, and $\epsilon_M = 1$, the quantity $\{A \, \beta \, \epsilon_M \}$ is in fact close to 3.  However, if we were to require that the $k$-range be very large, then $\beta$ would be larger, and $\epsilon_M$ would be required to be smaller than unity accordingly.

Inserting this parameter dependence into eq.(\ref{dEdt}) then leads back, to within the approximation that $A \, \beta  \, \epsilon_M = 3$, to the result we were seeking, 
\be
\frac{3}{2} \, M_{BH} \, c^2 \, H(z) \, {\left(\frac{1+z_{\star}}{1+z}\right)}^3 \; .
\label{balanceeq}
\ee
After integrating we obtain with this redshift dependence a constant dark energy density as in eq.( \ref{blackholeeq}) by multiplying by the redshift evolution of black holes $N_{BH,0} \, (1+z)^3$.  The factor of 4 multiplying $\rho_{DE}$ in eq.\ref{Econservation} derives from the sum of dark energy density and pressure, so corresponds to $3 \, (\rho_{DE} + P_{DE}/c^2)$.  Therefore the rate of change of dark energy with time is $3 \, (\rho_{DE} + P_{DE}/c^2) \, H(t)$ and today

\be
3 \rho_{DE} H(z=0) \; = \; \frac{3}{2} \, M_{BH} \, c^2 \, H(z = 0) \, {\left({1+z_{\star}}\right)}^3 \; .
\label{balanceeq2}
\ee

This shows that dark energy remains at the level of eq.(\ref{blackholeeq}) throughout the evolution of the universe, in the approximation that most early super-massive black holes were formed over a short span of time.

\subsection{Consistency}

Integrating the energy rate in eq.(\ref{dEdt}) over the time, using as the time interval the collapse time of the gravitating matter and using the inverse to three times the collapse time in eq.(\ref{kap}) instead of the Hubble parameter, the energy transferred is 
\bea
E\,=\, \frac{1}{2}\,M_{BH}\,c^2 \, ,
\eea
which is the same as the energy of the gravitational waves in eq.(\ref{InjEq}).  This does not prove the factor $1/2$, but serves as a plausibility check.
\par
Although the total mass of stellar-mass black holes in a galaxy is larger than that of the super-massive black hole at the center of most galaxies, stellar-mass black holes do not contribute appreciably to the production of gravitational waves through stimulated emission.  The ratio of the two phase space densities $\mathcal{N}$ and $g_b$ on the branes as calculated in our frame of reference
\be 
\frac{\mathcal{N}}{g_b} \, \simeq\,\frac{(8\,\pi)^2}{3}\,\left(\frac{M_{BH}}{m_{pl}}\right)^4\left(\frac{1+z_*}{1+z}\right)^4\,\frac{H(z)\,l_{pl}^3}{r^2(z,z_*)\,c}
\label{pvol}
\ee
must exceed $1$. The energy transfer occurs as long as $\mathcal{N}> g_b$, for the threshold function $ 2 \, g_b\,\mathcal{N}/(g_b+\mathcal{N})$, describing the interaction of gravitons. Using a value of $z_{\star} \,= \, 50\,$ in eq.(\ref{pvol}) we find a value of $10^{1.9}$ at which stimulated emission occurs.  Here the distance integral $r(z, z_{\star})$ is defined by ${\int_z}^{z_{\star}} (c \, d z)/H(z)$.  Since both $\mathcal{N}(k,t)$ and $g_b(p,t)$ are taken to be Planck distributions, emission occurs at all $k \, < \, k_{peak}$ simultaneously.  This condition has to hold for all the time since creation of the black holes:  So going through the times from $z_{\star}$ to some redshift $z$ below that, the condition has a minimum, which is about two orders of magnitude below today's value. This means that lower mass black holes lost their connection to the background at some early redshift.  Thus in the current epoch only black holes of  $\simge 10^6$ solar masses contribute to the stimulation process.

\subsection{Caveats}


First, we have used the assumption that there is actually a single identifiable first generation of super-massive stars, which evolve into black holes, all of very similar mass.  However, as can be easily ascertained, any later deposition of gravitational wave energy is weakened relative to the first by the product of the density of those black holes $N_{BH, 0}$ and the redshift factor $(1 + z_{\star})^3$.  An alternate possible picture would be the formation of degenerate dark matter stars, which get eaten up from the inside by stellar mass black holes \cite{MunyanezaBiermann05,MunyanezaBiermann06} to produce super-massive black holes.  Such a picture would probably generate very much weaker gravitational waves.  A mass of order $3 \cdot 10^{6} \; {\rm M_{\odot}}$ \cite{WangBiermann98} allows the big mass black holes to form relatively early in the universe as observed; any original mass significantly less would not allow sufficient time.  With these assumptions the dark energy density has a dominant value which has been nearly constant since the beginning of the epoch of black hole formation. 

Second, we have assumed, that black holes reach their high masses observed more by merging than by accretion, since with such a model we obtain more readily the relatively high early black hole densities necessary. The electrodynamic emission of accreting black holes integrated over time tends to exceed what is implied by cosmological simulations \cite{CarameteBiermann10}. All the energy deposition ``visible" in the sum is accounted for by large scale structure formation.  In this picture most of the merging of black holes would happen in the periods directly after their formation, so at redshifts far beyond current observations.


\section{Predictions}

For each super-massive black hole formed we have a soliton of gravitational waves slowly gaining energy with cosmic time as given in eq.(\ref{balanceeq}).  Mergers of black holes add solitons similarly.  The ensemble of these solitons constitute dark energy.  An immediate consequence is that in this model there was no dark energy from this process at recombination.

First, dark energy should be detectable by pulsar timing (e.g. \cite{Kramer10}).  Our proposal says that dark energy arises from coherent bursts of gravitational waves produced by forming the first generation of super-massive black holes.  This implies that at the properly redshifted frequency of these black holes the gravitational wave background should reach a value close to unity relative to critical density, with a narrowed Planck spectrum, which is broadened by the evolution of the black hole merger and formation history with $\nu_0$ extending to the low frequency side.  

Second, the passing of the soliton-like gravitational wave shells should be detectable as jitter in clocks (e.g. \cite{Predehletal12}), and in laser-interferometry.

Third, the original creation of the super-massive black holes should be detectable through their extended remnants in the radio, X-ray, $\gamma$-ray and neutrino background, and in ionized molecular hydrogen absorption.  It may have been detected already in the radio and neutrino backgrounds \cite{Fixsenetal11,Kogutetal11,Seiffertetal11,Condonetal12,ice}. We discuss this point in more depth elsewhere \cite{Zeuthen12}.



\section{Conclusions}

In this paper we have presented a simple analytic picture for the origin of dark energy, based on the idea that in their formation super-massive black holes eject a coherent burst of gravitational waves, which get about $(1/2) M_{BH} c^2$ from a background brane by stimulated emission.  This coherent burst continues to feed on the background by stimulated emission, and thus compensates the energy loss implicit in cosmic expansion, mimicking an equation of state $P_{DE} = - \rho_{DE} c^2$, obtained using the traditional Friedmann equations without energy transfer.  With the equations including energy transfer,   eqs.(\ref{FRW1}) and (\ref{FRW2}) the equation of state becomes $P_{DE}\,=\,\rho_{DE}\,c^2/3$, which is consistent with gravitational waves.   The background is conceived as similar to the model by Randall \& Sundrum \cite{RSa,RSb}, but variable in time.  We have used only the Einstein and conservation equations in 5D, require FRW to hold on both branes, take dark energy to be gravitational waves, or equivalently a gas of gravitons, and added stimulated emission to transfer energy from the strong brane, the background to our world.  This concept may give the correct energy density of dark energy, its time evolution, and also predicts that dark energy was weaker in the past (\cite{Planck2013XVI}).

There are several possible observational tests of these ideas, among them i) pulsar timing to detect the gravitational wave background which constitutes dark energy; ii) a jitter in clock timing associated with the many coherent bursts of gravitational waves going past; and, iii) radio and  neutrino backgrounds associated with the remnants of the formation of the first generation of black holes. 
\\

\section{Acknowledgements}

Discussions with L. Clavelli greatly contributed to the development of these ideas; discussions with L. Caramete and O. Micu (Bucharest, Romania), R. Casadio (Bologna, Italy), M. Cavaglia (Oxford, MS, USA), L. Gergely (Szeged, Hungary), Sh. Hou and A. Stern (Tuscaloosa, AL), P. Joshi (Mumbai, India), Gopal-Krishna (Pune, India), B. Nath and C. Sivaram (Bangalore, India), P. Nicolini (Frankfurt, Germany), and D. Stoikovic (Buffalo, NY, USA) are gratefully acknowledged.  
This research was supported in part by the DOE under grant DE-FG02-10ER41714.

\end{document}